\documentclass{article}

\usepackage{PRIMEarxiv}

\usepackage[utf8]{inputenc} % allow utf-8 input
\usepackage[T1]{fontenc}    % use 8-bit T1 fonts
\usepackage{hyperref}       % hyperlinks
\usepackage{url}            % simple URL typesetting
\usepackage{booktabs}       % professional-quality tables
\usepackage{amsfonts}       % blackboard math symbols
\usepackage{nicefrac}       % compact symbols for 1/2, etc.
\usepackage{microtype}      % microtypography
\usepackage{lipsum}
\usepackage{fancyhdr}       % header
\usepackage{graphicx}       % graphics
\usepackage{algpseudocode}
\usepackage{algorithm}
\usepackage{float}
\graphicspath{{media/}}     % organize your images and other figures under media/ folder

\title{Simulating Episodic Like Memory in Octopus: Breaking the Rules}
\author{
  Jeffrey L. Krichmar \\
  Department of Cognitive Sciences \\
  Department of Computer Science \\
  University of California, Irvine \\
  Irvine, CA 92697-5100, USA\\
  \texttt{jkrichma@uci.edu} \\
}

\begin{document}
\maketitle

\begin{abstract}
Episodic memory, which involves the recollection of past events, has been observed in several species including the invertebrate cuttlefish. In a subsequent experiment, episodic memory was not observed in the closely related octopus. Rather the octopus used simpler, cognitive rule-based strategies to gain access to food. Using a prior model of cuttlefish episodic-like memory, we replicated the octopus experiment. Under less challenging conditions, episodic memory and the simpler cognitive strategies all had positive gains. However, under more challenging conditions, which introduced the threat of predation, only the episodic memory model had positive gains. These results suggest that if challenged ecologically, it would be advantageous for octopus to use episodic-like memory. We discuss possible reasons for the behavioral discrepancies in these closely related cephalopods. 
\end{abstract}

% keywords can be removed

\section{Introduction}

Although episodic-like memory has now been observed in a range of organisms, it remains difficult to confirm the deployment of this cognitive function in non-human animals \cite{Davies2024}. Episodic memory involves the recollection of a personal experience. Recalling these experiences requires mental time travel to reflect on the past  \cite{Tulving2002}. Several species have demonstrated episodic-like memory behaviorally by recalling the ``what'', ``when'', and ``where'' of past events. For example, it has been shown that scrub jays remember when food items are stored by allowing them to recover preferred perishable and non-perishable food  \cite{clayton1998}. Scrub jays searched preferentially for fresh food if not much time passed. But if enough time had passed that the preferred food had decayed, they searched for the non-perishable food. These behavioral experiments demonstrated that corvids could recall the ``what'' (food type), ``when'' (short vs. long delay), and ``where'' (cache location) of their memory.  

Cuttlefish, an invertebrate cephalopod, have shown episodic-like memory in experiments similar to those carried out with scrub jays  \cite{JosetAlves2013, Schnell2021a}. In the first phase of these experiments, cuttlefish were trained to remember the locations of a preferred food (shrimp) and a non-preferred food (crabs).  In the second phase of the experiment, cuttlefish learned that the preferred food was available after a longer delay, and the non-preferred was the only food option available after a shorter delay. Like the scrub jays, cuttlefish recalled that shrimp (what), were available after a long delay (when), at a specific location (where). 

However, in a similar experiment with octopus, a closely related cephalopod to cuttlefish, strategies utilizing episodic memory were not observed  \cite{Poncet2022}. Instead the octopus favored simpler cognitive rule-based strategies such as alternation, familiarity, risk sensitivity, and win-stay.  In their experiment, octopuses were first presented with two pots that contained a preferred and less-preferred food. They were allowed to eat both. Food was then presented after a one or three hour delay. If one hour had passed, only the less preferred food was presented. If three hours passed, both food items were presented. In Poncet and colleagues' experiments, six out of seven octopuses relied on less-cognitively demanding strategies than keeping track of the time when preferred food was available. It is not clear if this was due to experimental design or different ecological demands of octopus compared to cuttlefish. 

Given that both anecdotally and experimentally, octopus have demonstrated flexibility in their decision-making, planning and reasoning  \cite{Birch2020,Hanlon2018,Schnell2019,Schnell2021b}, we wondered if there were reasons that a more complicated strategy, such as deploying episodic memory, was not observed in the octopus. It may have been that food was readily available even if was not the organism's preference, or the octopuses were not under environmental pressure due to predation or other threats and simple strategies sufficed.

In prior work, we created a model to simulate the episodic-like memory experiments carried out in the cuttlefish  \cite{Kandimalla2025}. The memory model, which was a three-dimensional matrix, indexed by what, when, and where tuples, was sufficient to replicate cuttlefish episodic-like memory experiments. To further challenge the model, we created a predator-prey scenario in which the simulated cuttlefish agent had to remember what item (i.e., predator, crab, shrimp) would appear when, and where it could find prey and avoid predators. 

We hypothesized that if this model was compared to the episodic-like experimental paradigm in \cite{Poncet2022}, that simple cognitive strategies might be favorable. However, these simple strategies may fall short with the introduction of an environmental threat, such as a predator.  We test these hypotheses with the model introduced in  \cite{Kandimalla2025}.

\section{Methods}
\label{sec:methods}

Two episodic memory scenarios were developed. The first modeled experiments that demonstrated episodic-like memory experiments \cite{JosetAlves2013,Poncet2022}.  The second involved a predator-prey scenario. The details for the episodic memory model were originally described in  \cite{Kandimalla2025}. We reiterate many of those details here for completeness. The source code for both these simulations is written in Python and will be made available upon publication in a peer-reviewed journal. 

\subsection{Episodic Memory Model}
\label{sec:EpiMemModel}

At the core of both scenarios is a three-dimensional memory model that is indexed by ``what'', ``when'', or ``where'', as well as combinations of these indices. The structure holds the expected values of these indices. For example, a query of ``what[shrimp]'' and ``when[hour 3]'' would return the expected values of shrimp at the 3\textsuperscript{rd} hour over all locations in the environment (see Equation \ref{eqn:value}). 

\begin{equation}
   \vec{v} \gets epimem[shrimp, 3, :]
   \label{eqn:value}
\end{equation}

The result from this query is the vector $\vec{v}$ that contains the expected value of shrimp at all the ``where'' locations in memory.  This vector can then be used by a reinforcement learning algorithm to select the appropriate action (e.g., hunting or roaming), with an exact action set depending on the scenario described below.

We use the common \textit{delta rule} to learn associations of what, when, and where. More sophisticated learning rules could be applied but the delta rule will suffice for these simulations to demonstrate episodic-like memory. The learning rule is given by Equation \ref{eqn:delta}.
\begin{equation}
   \Delta epimem[what, when, where] = \alpha (r - epimem[what, when, where]), 
   \label{eqn:delta}
\end{equation}
where $\alpha$ is the learning rate and $r$ is the reward or penalty incurred. 

For example, if a shrimp was found and eaten at the 3\textsuperscript{rd} hour at location 42 and received a reward of 4 points with learning rate of 0.10, then Equation \ref{eqn:update} would be:
\begin{equation}
   epimem[shrimp, 3, 42] \gets epimem[shrimp, 3, 42] + 0.10(4 - epimem[shrimp, 3, 42])
   \label{eqn:update}
\end{equation}

Once a value vector $\vec{v}$ is acquired for a given action, we subject it to the Softmax function. In the case of equation \ref{eqn:value}, the vector would be a list of expected values at each location and the action $act$ given by Equation \ref{eqn:softmax} would be go to a specific location.

\begin{equation}
    act = \frac{\beta e^{v_i}}{\Sigma_j{\beta e^{v_j}}}
    \label{eqn:softmax}
\end{equation}

Combinations of memory queries can be used to obtain expected values and agent actions. For example, one could query ``what'' and ``when'' for each object (e.g., crab, shrimp, predator), then subject these values to equation \ref{eqn:softmax}, which may choose \textit{hunt shrimp} as the best action at this time. Then an additional query of ``shrimp'' and ``current time'' would give the expected location of shrimp at the present time.

\subsection{Alternative Foraging Strategies}
\label{sec:AltStrat}

Poncet et al. observed the octopus using foraging strategies that were simpler than utilizing episodic memory for gathering food \cite{Poncet2022}. They termed these strategies: \textit{Familiarity}, \textit{Risk Sensitivity}, \textit{Alternation}, and \textit{Win-Stay}.  Each of these  strategies were coded in the simulation as alternatives to the episodic memory model.

\begin{algorithm}[H]
\caption{\textbf{Familiarity.} Octopus chose the same food pot on the second presentation as in the first presentation.}
\label{alg:fam}
\begin{algorithmic}
\If{$choice(1) = crab$} 
    \State $choice(2) \gets crab$
\Else
    \State $choice(2) \gets crab$
\EndIf 
\end{algorithmic}
\end{algorithm}

\begin{algorithm}[H]
\caption{\textbf{Risk Sensitivity.} Octopus chose the always available, but less preferred food source (i.e., crab in the present simulations)}
\label{alg:risk}
\begin{algorithmic}
\State $choice \gets crab$
\end{algorithmic}
\end{algorithm}

\begin{algorithm}[H]
\caption{\textbf{Alternation.} Octopus alternated their choice between trials $t$ and $t+1$.}
\label{alg:alt}
\begin{algorithmic}
\If{$choice(t) = crab$} 
    \State $choice(t+1) \gets shrimp$
\Else
    \State $choice(t+1) \gets crab$
\EndIf 
\end{algorithmic}
\end{algorithm}

\begin{algorithm}[H]
\caption{\textbf{Win-Stay.} If octopus were rewarded with the best food option in trial $t$, they maintained their choice in trial $t+1$. Otherwise, they shifted their choice to the other food option in trial $t+1$.}
\label{alg:win}
\begin{algorithmic}
\If{$reward(t) > 0$}
    \If{$choice(t) = crab$}
        \State $choice(t+1) \gets crab$
    \Else
        \State $choice(t+1) \gets shrimp$
    \EndIf
\Else
    \If{$choice(t) = crab$}
        \State $choice(t+1) \gets shrimp$
    \Else
        \State $choice(t+1) \gets crab$
    \EndIf
\EndIf 
\end{algorithmic}
\end{algorithm}

\subsection{Simulation Environments}
\label{sec:simenv}

We describe two different environments to test the episodic memory model. In both cases, we used a grid world containing an octopus agent, shrimp, and crabs. Predators were introduced in the Predator-Prey Scenario. Predators were assumed to detect the octopus at a greater distance, while the octopus's vision allowed it to only perceive nearby objects.

\subsubsection{Episodic Like Memory Scenario}
\label{sec:episcenario}

\begin{figure}[ht]
    \centering
    \includegraphics[width=0.67\linewidth]{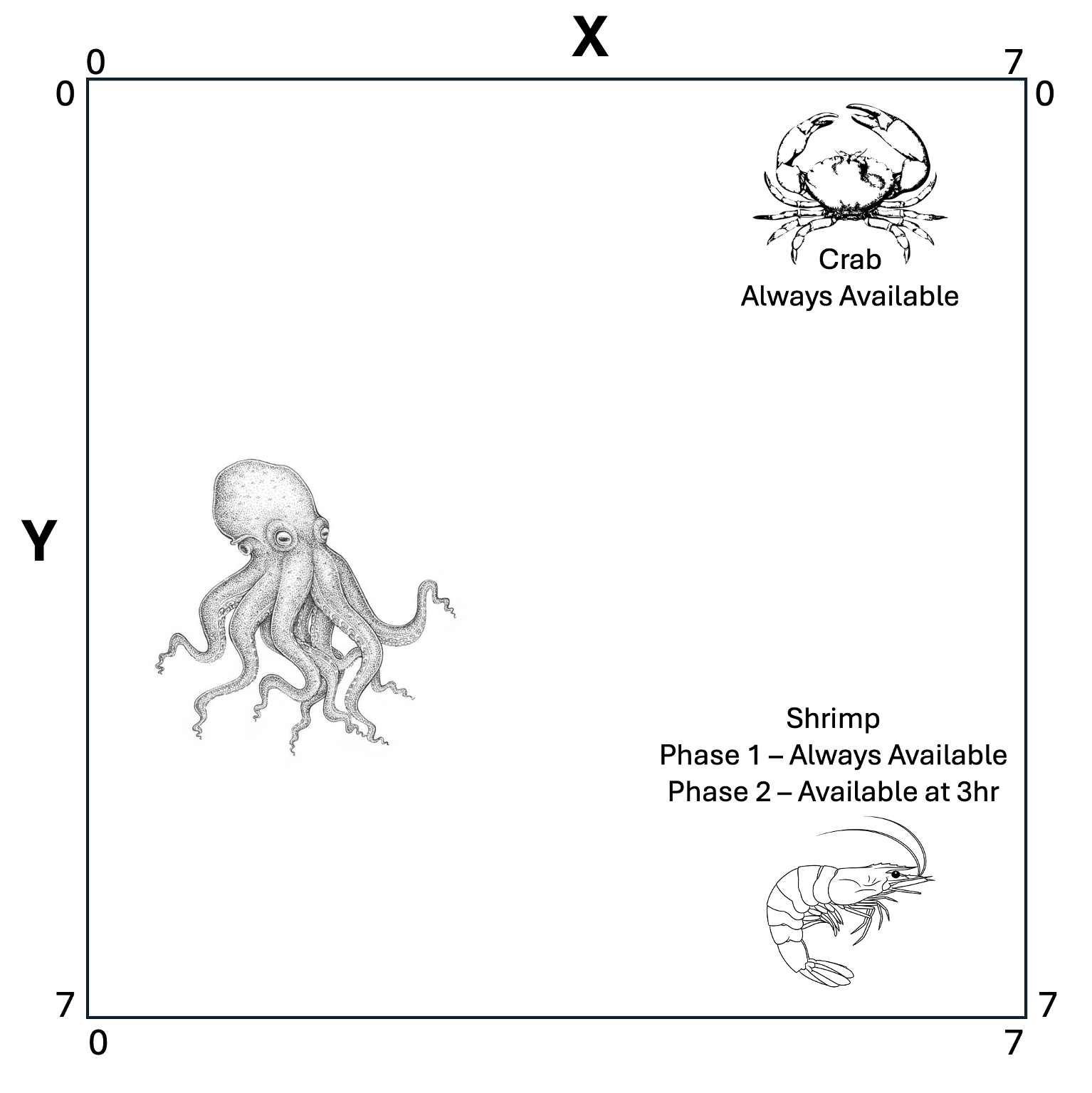}
    \caption{\textbf{Episodic-Like Memory Environment}. At the start of each hour, the octopus agent is placed in the middle of the left side of the environment (x=0, y=4), and a crab is placed in the upper right region (x=7, y=0). The crab is available every hour. The shrimp's location is in the bottom right region (x=7,y=7). During phase 1, the shrimp is available every hour. During phase 2, the shrimp is only available during hour 3. The octopus agent can move freely within the environment, shrimp and crab remain stationary.}
    \label{fig:epi_like_env}
\end{figure}

Episodic memory was investigated in a two phase experiment. In the first phase, the octopus agent learned where crab and shrimp were located. In the second phase, the octopus agent learned that shrimp, which is a preferred food, was only available after a 3 hour delay. To simulate this experiment, an environment was created with two objects, a crab and a shrimp, a 3 hour duration, and an 8x8 grid world environment (see Figure \ref{fig:epi_like_env}).  This created an episodic memory that was initialized to:

\begin{equation}
    epimem[what_0..what_1, when_0..when_2, where_0..where_{63}] = 0.0
    \label{eqn:epilikemem}
\end{equation}

\begin{table}[ht]
    \centering
    \caption{Episodic Like Memory Scenario Parameters}
    \begin{tabular}{|c|c|}
        \hline
        $\alpha$ & $0.10$ \\
        \hline
        $\beta$ & $1.0$ \\
        \hline
        Octopus Initial Position (x,y) & $(0,4)$ \\
        \hline
        Octopus Vision & $< 2$ \\
        \hline
        Crab Reward & [$1.0$ | $2.0$] \\
        \hline
        Crab Position (x,y) & $(7,0)$ \\
        \hline
        Shrimp Reward & $4.0$ \\
        \hline
        Shrimp Position (x,y) & $(7,7)$ \\
        \hline
        Roam Value & $0.5$ \\
        \hline
    \end{tabular}
    \label{tab:epiparms}
\end{table}

As in  \cite{JosetAlves2013}, both the crab and shrimp are available during all 3 hours in the first phase of the experiment. In the second phase, the crab is always available, but the shrimp location is only rewarded after 3 hours. In both phases, if the agent reaches an available food item, the appropriate reward (see Table \ref{tab:epiparms}) is applied to the memory as given by Equation \ref{eqn:delta}. 

The simulation ran for 200 days, where each day lasted 3 hours, and an hour lasted 100 simulation time steps. The three actions were: \textit{hunt crab}, \textit{hunt shrimp}, and \textit{roam}. The value for \textit{roam} was set to 0.5, the other action values were learned by the model (see Equation \ref{eqn:delta}). At the start of every hour: 1) Values were acquired with query for each object (crab, shrimp) at the current time (see Equation \ref{eqn:value}). For each object, the maximum value across all locations was chosen. 2) An action was chosen with the Softmax function (see Equation \ref{eqn:softmax}). The average value or a sum of values could be used, as will be seen in the other scenario. 3) If the action is to \textit{hunt}, an additional query then retrieves the location where the chosen object has its maximum value. It is assumed that the octopus agent knows how to reach this location. 4) If the action is to \textit{roam}, then the octopus agent moves randomly a step in one of 8 cardinal directions. 5) For all 3 actions, if a shrimp or crab is available at this time, and the octopus agent is less than 2 steps from the object, the value is updated by Equation \ref{eqn:delta}. The parameters for this scenario are given in Table \ref{tab:epiparms}.

\subsubsection{Predator-Prey Scenario}
\label{sec:predpreyscenario}

\begin{figure}[ht]
    \centering
    \includegraphics[width=0.67\linewidth]{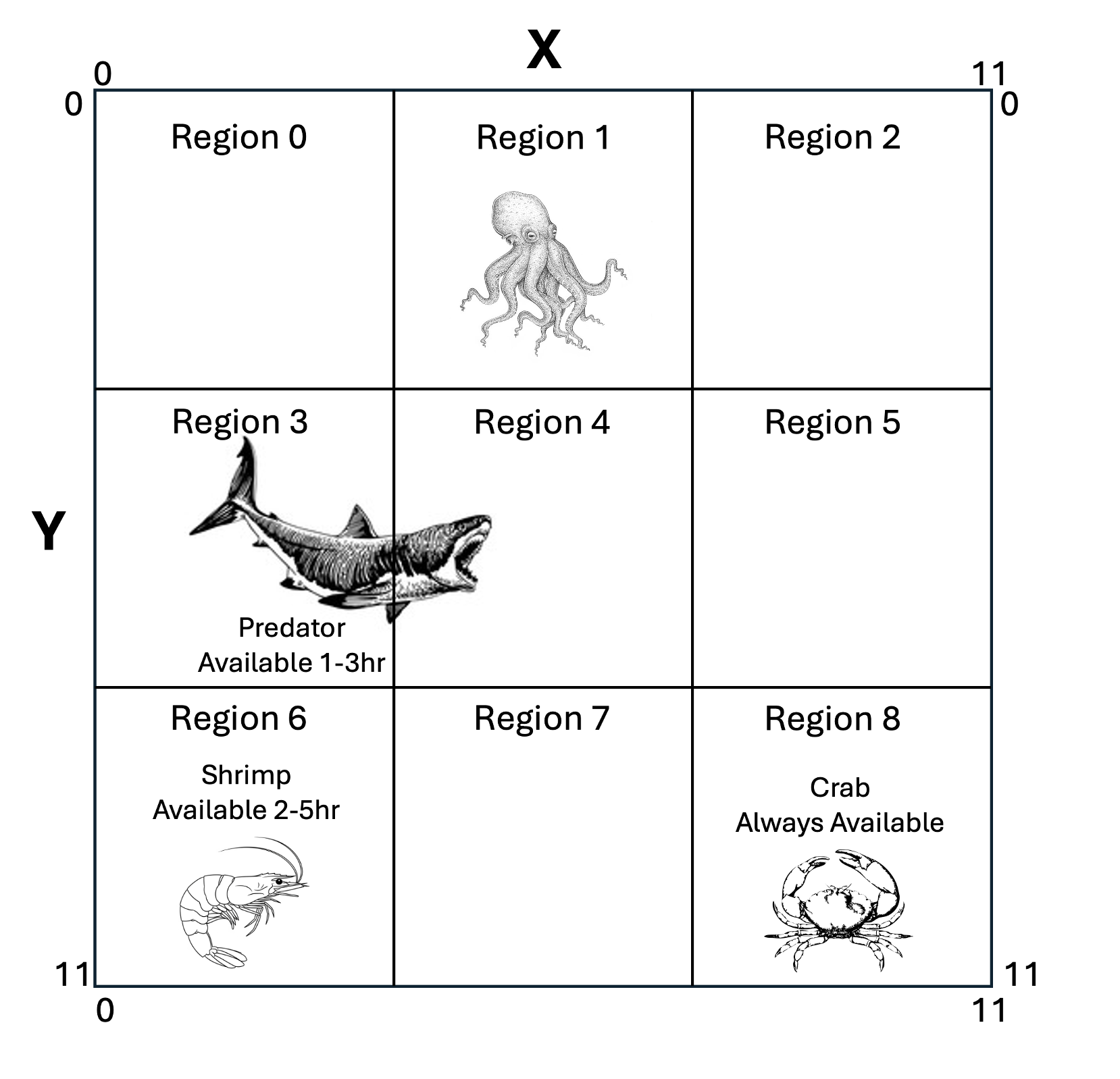}
    \caption{\textbf{Predator-Prey Environment}. The simulation environment is divided into 9 regions. Every hour, the octopus agent is placed at the top (x=6, y=0) of region 1 and can move freely between regions. Shrimp are available from hour 2 to hour 5 in region 6. Crab are available all day (hour 0 to hour 5) in region 8. The predator is in the environment from hour 1 to hour 3. Every hour when available, the shrimp and crab are placed anywhere in their respective regions and stay stationary. Every hour when present, the predator is placed somewhere in the left side (X < 6) of the environment.  The predator either moves randomly or approaches a octopus if it is within the predator's vision.}
    \label{fig:predprev_env}
\end{figure}

To further test our episodic memory model, we created a predator-prey scenario in which the octopus agent learned when and where shrimp and crab were likely to be located, but also needed to learn when and where a predator could be located. To simulate this experiment, an environment was created with three objects, a crab, a shrimp, and a predator, a 6 hour duration, and a 12x12 grid (see Figure \ref{fig:predprev_env}). The parameters for this scenario are given in Table \ref{tab:predpreyparms}.  This created an episodic memory that was initialized to:

\begin{equation}
   epimem[what_0..what_2, when_0..when_5, where_0..where_{143}] = 0.0
\end{equation}

\begin{table}[ht]
    \centering
    \caption{Predator-Prey Scenario Parameters}
    \begin{tabular}{|c|c|}
        \hline
        $\alpha$ & $0.10$ \\
        \hline
        $\beta$ & $1.0$ \\
        \hline
        Octopus Initial Position (x,y) & $(6,0)$ \\
        \hline
        Octopus Vision & $< 2$ \\
        \hline
        Predator Initial Position (x,y) & $(0..6,0..11)$ \\
        \hline
        Predator Vision & $< 4$ \\
        \hline
        Predator Penalty & $8$ \\
        \hline
        Crab Reward & $1.0$ \\
        \hline
        Crab Region & $8$ \\
        \hline
        Shrimp Reward & $4.0$ \\
        \hline
        Shrimp Region & $6$ \\
        \hline
        Roam Value & $0.5$ \\
        \hline
    \end{tabular}
    \label{tab:predpreyparms}
\end{table}

The simulation ran 200 days, where each day lasted 6 hours, and an hour lasted 100 simulation time steps. The four actions were: \textit{hunt crab}, \textit{hunt shrimp}, \textit{hide}, and \textit{roam}. The value for \textit{roam} was set to 0.5, the other action values were learned by the model (see Equation \ref{eqn:delta}). At the start of each hour, objects were placed in the appropriate regions of the environment (see Figure \ref{fig:predprev_env}). If the action was \textit{hunt}, the octopus agent headed to the center of region where the prey was most likely to be found. Once in the region, the agent randomly roamed the region until a prey was found, whereby it receives the appropriate reward, or the hour had passed. If the action was \textit{hide}, the agent simulated camouflaging. That is, the octopus was invisible to the predator.  The predator randomly roamed the environment until it saw the octopus was within a distance of 4. In that case, the predator would head directly toward the octopus.

At the start of every hour: 1) Values were acquired with a ``what'' query for each object (crab, shrimp) at the current time (see Equation \ref{eqn:value}). In this experiment, the total value summed over all locations for each object. 2) An action was chosen with the Softmax function (see Equation \ref{eqn:softmax}). 3) If the action was to \textit{hunt}, an additional query then retrieves the region where the chosen object has its maximum value. The agent then proceeded to the center of the selected region. It is assumed that the octopus agent knows how to reach this location. 4) If the action is to \textit{roam}, then the octopus agent moves a step in one of 8 cardinal directions. The direction is chosen randomly. 5) For all 3 actions except \textit{hide}, if a shrimp or crab are available at this time, and the octopus agent is less than 2 steps from the object, the value is updated by Equation \ref{eqn:delta}.

\section{Results}
\label{sec:results}

The episodic memory model of Section \ref{sec:EpiMemModel} was compared to the alternative foraging strategies described in \ref{sec:AltStrat} for both the episodic-like memory and predator-prey scenarios.

\subsection{Episodic-Like Memory Experiments}
\label{sec:epilikeexp}

\begin{figure}
    \centering
    \includegraphics[width=0.75\linewidth]{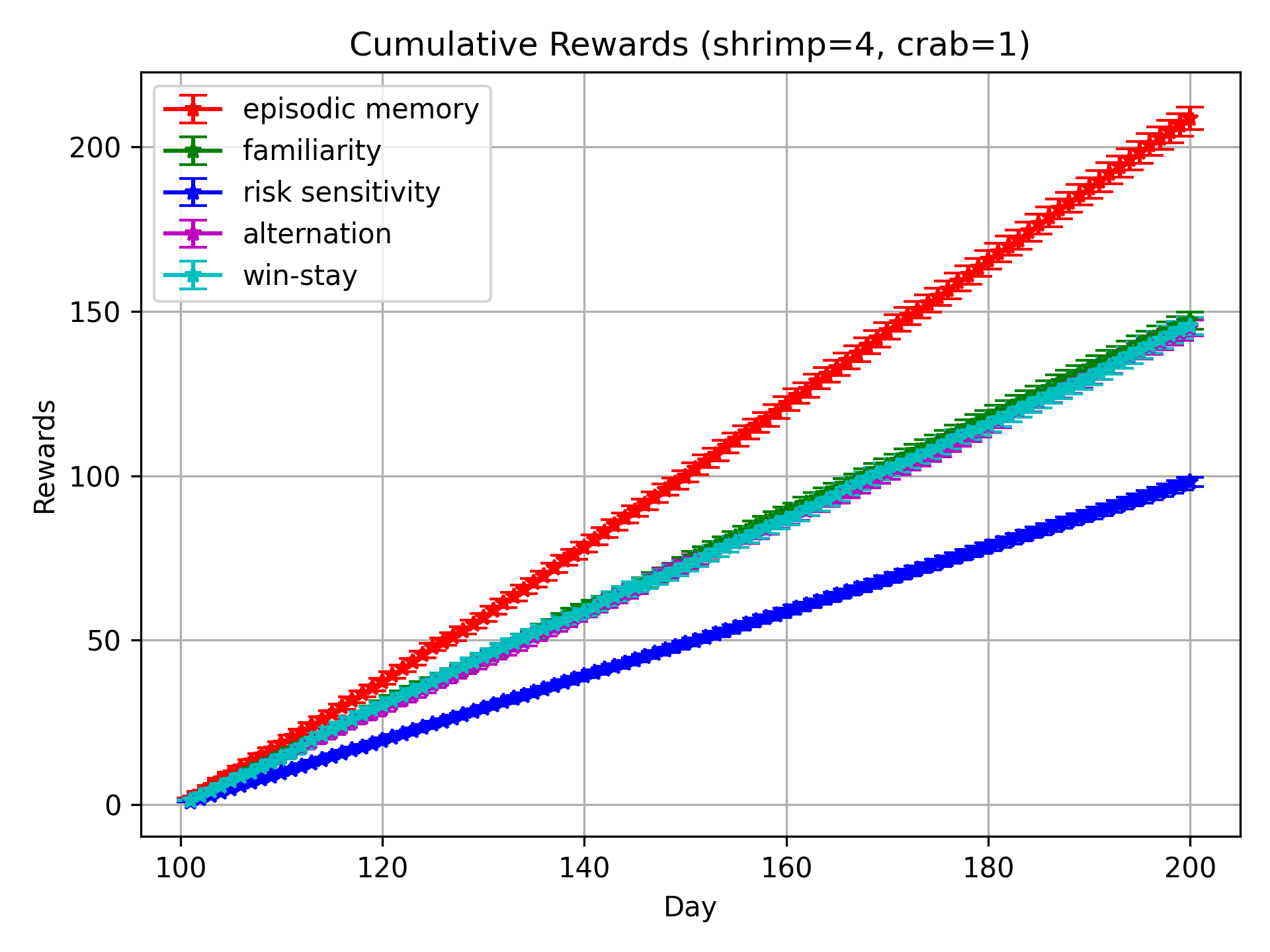}
    \caption{\textbf{Cumulative Rewards in the Episodic-Like Memory Environment}. Although the Episodic Memory model had the best performance, the alternative foraging strategies still showed positive gains. The preferred shrimp reward was $4$ and the less preferred crab reward was $1$. The mean cumulative awards for all the strategies are shown for the last $100$ days of the simulation. The error bars denote the standard error of the mean.}
    \label{fig:epi_like_cum_4_1}
\end{figure}

In the episodic-like memory experiments, which followed the experiments in the cuttlefish \cite{JosetAlves2013} and octopus \cite{Poncet2022}, both the episodic memory model and the alternative foraging strategies showed positive gains (see Figure \ref{fig:epi_like_cum_4_1}. Even though the episodic memory model had the best performance, the agents using alternative strategies still were getting rewards. This implies that these simple strategies, which take less cognitive effort, are advantageous.

\begin{figure}
    \centering
    \includegraphics[width=1.0\linewidth]{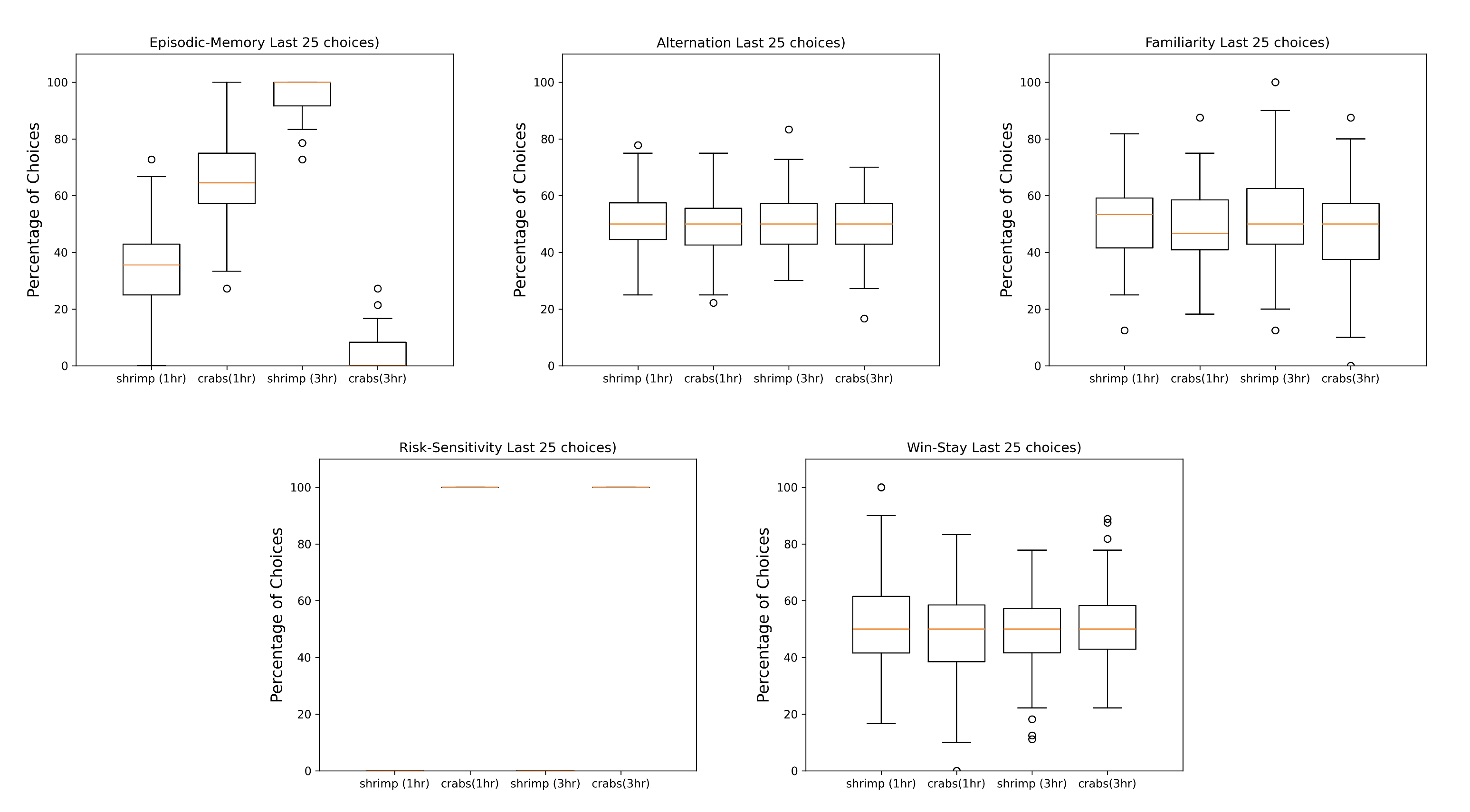}
    \caption{\textbf{The percentage choices during the last $25$ trials are shown for all strategies.} Only the Episodic Memory model could recognize that $3$ hour trials had different reward policies. The Risk Sensitivity Strategy always chose crab and therefore had no variation. The red line in the boxplot is the median of the $100$ simulation runs, the box extends from the first quartile to the third quartile, the whiskers extend $1.5x$ beyond the quartiles. Circles denote outliers.}
    \label{fig:epi_like_strats}
\end{figure}

Only the episodic memory model learned to take the passage of time into consideration.  Figure \ref{fig:epi_like_strats} show the choices made by the agents for each strategy.  The episodic memory model tended to choose crabs if one hour had passed and shrimp if three hours had passed. The risk sensitivity strategy always chose crab by definition. The other strategies tended toward random choice, since the delay between food presentations was randomized.

\begin{figure}
    \centering
    \includegraphics[width=0.75\linewidth]{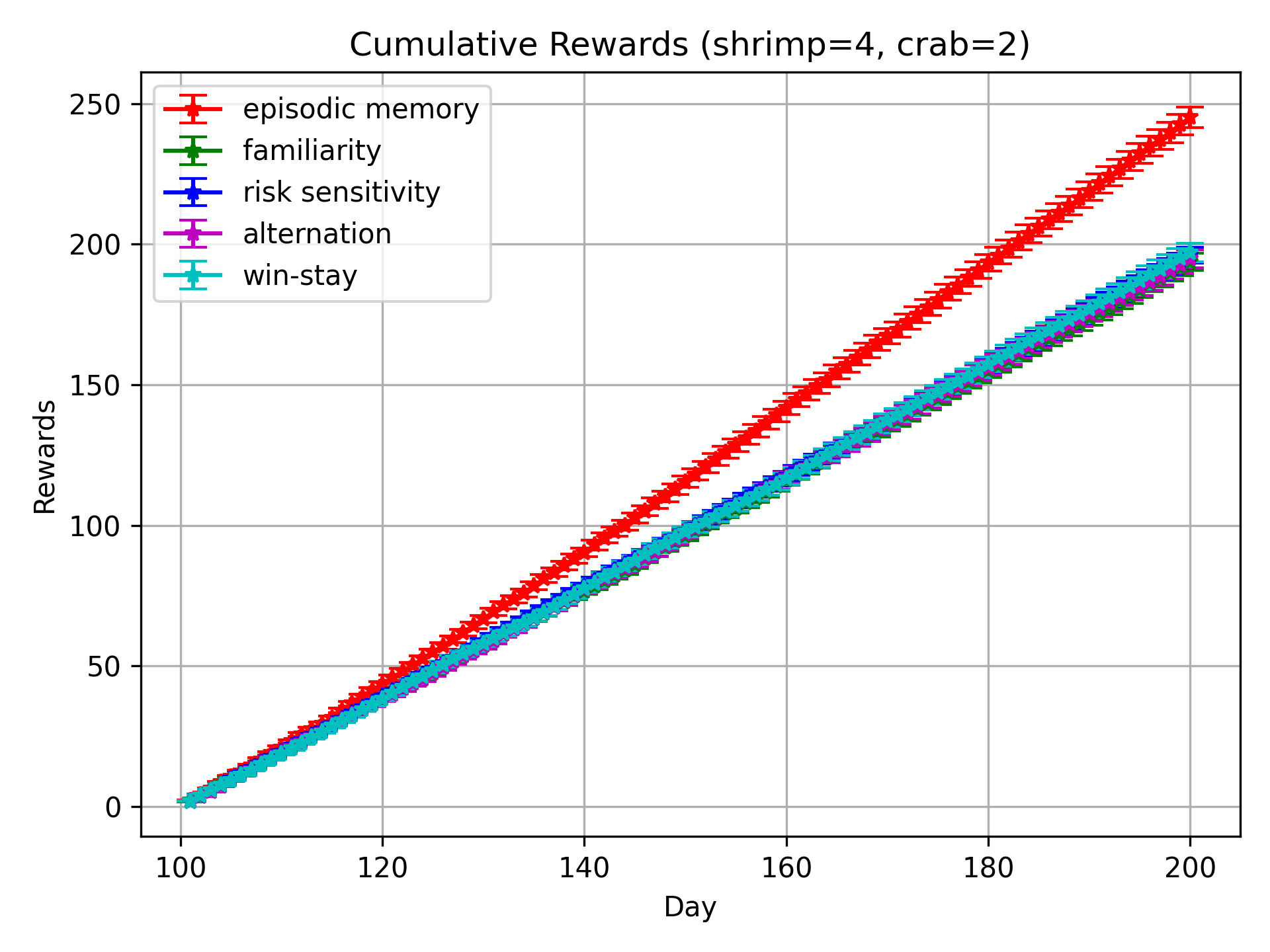}
    \caption{\textbf{Cumulative Rewards in the Episodic-Like Memory Environment when the crab food option is worth more}. When the less preferred crab reward was worth $2$, the performance of the alternative strategies was closer to the episodic memory model. The mean cumulative awards for all the strategies are shown for the last $100$ days of the simulation. The error bars denote the standard error of the mean.}
    \label{fig:epi_like_cum_4_2}
\end{figure}

The difference between strategies depends on the organisms food preferences and its hunger level. In both the cuttlefish and the octopus, there was no way of knowing how much one food was preferred over another.  The simulations so far assumed the shrimp was preferred four times more than the crab. However, if the shrimp were twice as preferred over crab, the simpler strategies become even more advantageous as compared to the episodic memory strategy (see Figure \ref{fig:epi_like_cum_4_2}). This further suggests that deploying simpler strategies may depend on the organism's preferences.

\subsection{Predator-Prey Experiments}
\label{sec:predpreyexp}

The episodic-like experiments simulated in Section \ref{sec:epilikeexp} and in the cephalopods were low risk. There was no threat of predation and food was always available even if it might not have been the organism's first choice. In  \cite{Kandimalla2025}, a predator-prey scenario was introduced to further challenge the episodic memory model.  We asked how the simpler strategies might fare in this more challenging environment.

\begin{figure}
    \centering
    \includegraphics[width=0.75\linewidth]{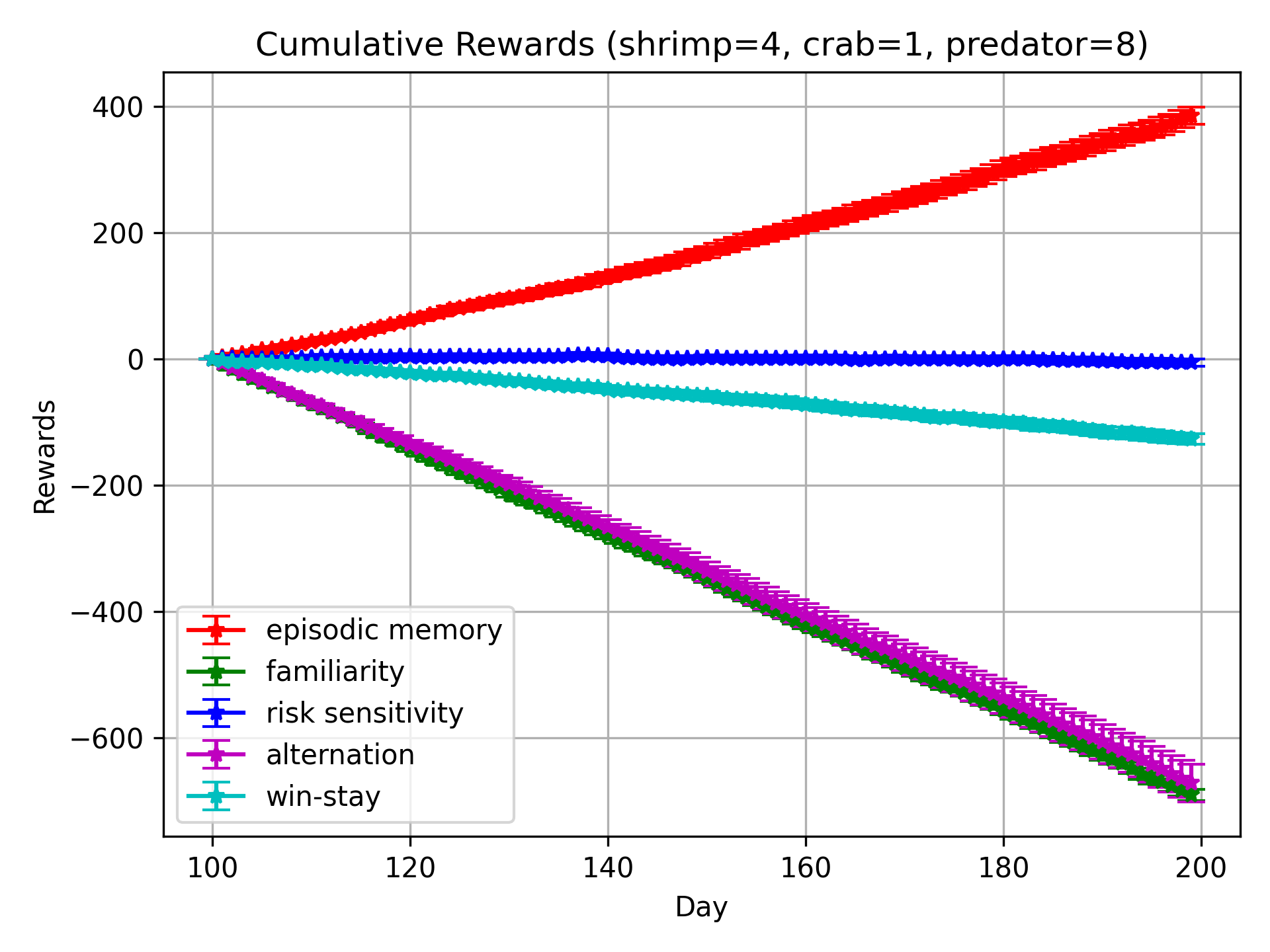}
    \caption{\textbf{Cumulative Rewards in the Predator-Prey Environment}. The Episodic Memory model was the only strategy that showed positive gains in rewards. The preferred shrimp reward was $4$, the less preferred crab reward was $1$, and getting caught by a predator had a penalty of $-8$. The error bars represent the standard error of the mean. The mean cumulative awards for all the strategies are shown for the last 100 days of the simulation.}
    \label{fig:predprey_cum_awards}
\end{figure}

When subjected to the threat of predation, only the episodic memory model had positive gains (see Figure \ref{fig:predprey_cum_awards}). This is because the model learned that a shark (\textit{what}) was present (\textit{where}) at certain times (\textit{when}). The simpler strategies did not have this knowledge and were often ``eaten'' by the shark. Thus, having negative cumulative rewards. The risk sensitivity strategy tended toward zero because in always choosing crabs the octopus agent happened to be in locations further from the shark. If the crab position were moved or only available when the shark was present, the cumulative rewards for risk sensitivity would also be negative like other simple strategies.

Taken together, these simulation results suggest an explanation for why the octopus were not observed to deploy episodic memory. Octopus might make the effort to deploy their memory capabilities when in demanding situations. This and alternative explanations are discussed in the next section.

\section{Discussion}

Environmental pressures may dictate the deployment of different cognitive strategies and functions. Comparing a model of episodic memory to simpler cognitive strategies, we showed that under less challenging conditions, simpler foraging strategies, while not optimal, were still advantageous. That is, our agent did not go hungry. However, the introduction of environmental challenges in the form of a predator showed that deploying episodic memory had clear survival advantages.

An open question though is why was episodic memory observed in cuttlefish, but not octopus under similar conditions  \cite{JosetAlves2013,Poncet2022}. We suggest several reasons below. 

Cuttlefish and octopus evolved from the shelled \textit{mollusca} phylum and it has been hypothesized that the loss of this shell led to their intelligent behavior  \cite{Amodio2019}. Without a shell, these cephalopods had less protection from predators and needed to evolve evasive strategies (e.g., camouflage, ink clouds), as well as smart hunting strategies. This supports the \textit{Ecological Intelligence Hypothesis} in which complex cognition to exploit food resources, required spatial memory, decision-making, and inhibitory control \cite{Rosati2017}. The \textit{coleioda}, which is a subclass of \textit{cephalopoda}, includes cuttlefish, octopus and squid. Another cephalopod that still has a shell is the \textit{Nautilus}. Compared to other cephalopods, the nautilus has less sophisticated behavior and less complex nervous systems \cite{Hanlon2018,NixonYoung2003}.

It has been suggested that episodic memory in mammals requires a hippocampus  \cite{Tulving1998}. The  hippocampus receives  sensory input from numerous brain regions providing ``what'' information \cite{RollsTreves2024, Sanchez2021, TrevesRolls1992}. Hippocampal place cells demonstrate a neural correlate of ``where'' \cite{OKeefe1971}. Furthermore, place cells show both retrospective coding of past locations and prospective coding of future locations \cite{Ferbinteanu2003}, as well as encoding multiple timescales \cite{Buzsaki2026,Eichenbaum2014}. This suggests something like a ``when'' neural correlate in the hippocampus. 

Interestingly, the vertical lobe of cuttlefish and octopus, which is a critical site for learning and memory, have anatomical similarities to the mammalian hippocampus \cite{Shomrat2015, Young1991}. Like the hippocampus, there is fan-out multimodal input into a multilayered structure, and fan-in output that then projects to brain regions dictating behavior. Furthermore, like the hippocampus, the cephalopod vertical lobe supports long-term plasticity  \cite{Shomrat2011}.

Given the behavioral and anatomical similarities between cuttlefish and octopus, it would be surprising if they didn't both possess something like episodic memory. Poncet et al. suggested that that other cognitive strategies might have subsumed the need for episodic-like memory \cite{Poncet2022}. However, this does not explain why cuttlefish tended to use episodic-like memory in a similar situation. They also noted that availability of food might have lowered the motivation for learning the different food schedules. Comparing our results in the two scenarios also suggest that motivation might have been a driving factor in deploying episodic memory.

On the other hand, why were these simpler rule-based strategies not observed in cuttlefish in other experiments \cite{JosetAlves2013,Schnell2021a}? The discrepancy may be due to anatomical differences. Compared to the cuttlefish, which mainly uses its arms reach out and grab prey, octopus arms are far more complex having incredible sensitivity and degrees of freedom \cite{Kuuspalu2022,Nesher2020,Olson2024}. Moreover, unlike cuttlefish, which contain a cuttlebone for buoyancy, octopus bodies are more flexible and can fit into extremely small spaces. Anecdotally, this complexity has led to tales of octopus escaping enclosures in labs, restaurants and aquariums, and using coconut shells and other items to hide and ``walk''. Their incredible behavior in the wild has been well-documented on social media platforms. It might just be that octopuses have more embodied tricks up their sleeves than cuttlefish, which may require cuttlefish rely more on complex cognition to solve similar problems.

We should emphasize that these speculations need to be tested experimentally. Our simulations make testable predictions, but it will take clever experimental designs to confirm or reject these predictions. Nonetheless, the simulation of the behavior of these remarkably bright creatures further suggests that environmental pressures can lead to convergent cognitive complexity.

\section*{Acknowledgments}
This was was supported in part by National Institute of Neurological Disorders and Stroke award R01NS135850-02. The author thanks the Cognitive Anteater Robotics Laboratory (CARL) and the CuttleBot team for many productive discussions.

%Bibliography
\bibliographystyle{unsrt}  
\bibliography{references}

\end{document}